\documentclass[sigconf,table]{acmart}

\usepackage{booktabs} 
\usepackage{tabularx} 
\usepackage[english]{babel}
\usepackage{mathrsfs}

\usepackage{multirow}
\usepackage{multicol}
\usepackage{array}
\usepackage{arydshln}

\usepackage{amsmath}
\usepackage{amssymb}
\usepackage{booktabs}
\usepackage[inline]{enumitem}
\usepackage{siunitx}
\usepackage{stackengine}
\usepackage{extarrows}
\usepackage{tablefootnote}
\usepackage{caption}
\usepackage{subcaption}

\usepackage{footnote}
\makesavenoteenv{tabular}
\makesavenoteenv{table}

\usepackage{xspace} 

\newcolumntype{L}[1]{>{\raggedright\arraybackslash}p{#1}}
\newcolumntype{C}[1]{>{\centering\arraybackslash}p{#1}}
\newcolumntype{R}[1]{>{\raggedleft\arraybackslash}p{#1}}

\newcommand{\incl}{\emph{incl. }}

\copyrightyear{2020} 
\acmYear{2020} 
\acmConference[SIGIR '20]{}{July 25--30, 2020}{Xi'an, China}

\author{Sebastian Hofst\"{a}tter}
\affiliation{%
  \institution{TU Wien}
}
\email{s.hofstatter@tuwien.ac.at}

\author{Hamed Zamani}
\affiliation{%
  \institution{Microsoft}
}
\email{hazamani@microsoft.com}

\author{Bhaskar Mitra}
\affiliation{%
  \institution{Microsoft}
}
\email{bmitra@microsoft.com}

\author{Nick Craswell}
\affiliation{%
  \institution{Microsoft}
}
\email{nickcr@microsoft.com}

\author{Allan Hanbury}
\affiliation{%
  \institution{TU Wien}
}
\email{allan.hanbury@tuwien.ac.at}

\begin{document}

\title{Local Self-Attention over Long Text for Efficient Document Retrieval}

\begin{abstract}
Neural networks, particularly Transformer-based architectures, have achieved significant performance improvements on several retrieval benchmarks. When the items being retrieved are documents, the time and memory cost of employing Transformers over a full sequence of document terms can be prohibitive. A popular strategy involves considering only the first $n$ terms of the document. This can, however, result in a biased system that under retrieves longer documents. In this work, we propose a local self-attention which considers a moving window over the document terms and for each term attends only to other terms in the same window. This local attention incurs a fraction of the compute and memory cost of attention over the whole document. The windowed approach also leads to more compact packing of padded documents in minibatches resulting in additional savings. We also employ a learned saturation function and a two-staged pooling strategy to identify relevant regions of the document. The Transformer-Kernel pooling model with these changes can efficiently elicit relevance information from documents with thousands of tokens. We benchmark our proposed modifications on the document ranking task from the TREC 2019 Deep Learning track and observe significant improvements in retrieval quality as well as increased retrieval of longer documents at moderate increase in compute and memory costs.
\end{abstract}


\maketitle
\section{Introduction}
\label{sec:intro}

Deep neural networks have yielded dramatic improvements in several information retrieval (IR) tasks~\citep{guo2019survey,mitra2018introduction}.
Some of the improvements can be attributed to Transformer-based architectures~\citep{vaswani2017attention}, such as in BERT-based ranking models~\citep{nogueira2019passage, yan2020idst} and the Transformer-Kernel pooling (TK) model~\citep{Hofstaetter2020_ecai}.
These Transformer-based architectures employ self-attention to learn contextual embeddings of text for matching.
Unfortunately, the time and memory complexity of applying self-attention over a sequence of length $L$ is $O(L^2)$~\citep{kitaev2020reformer}.
When the goal is to retrieve documents, the cost of applying attention over whole documents can be prohibitive.

\begin{figure}
    \centering
    \includegraphics[width=0.48\textwidth,clip, trim=0.5cm 0.2cm 0cm 0cm ]{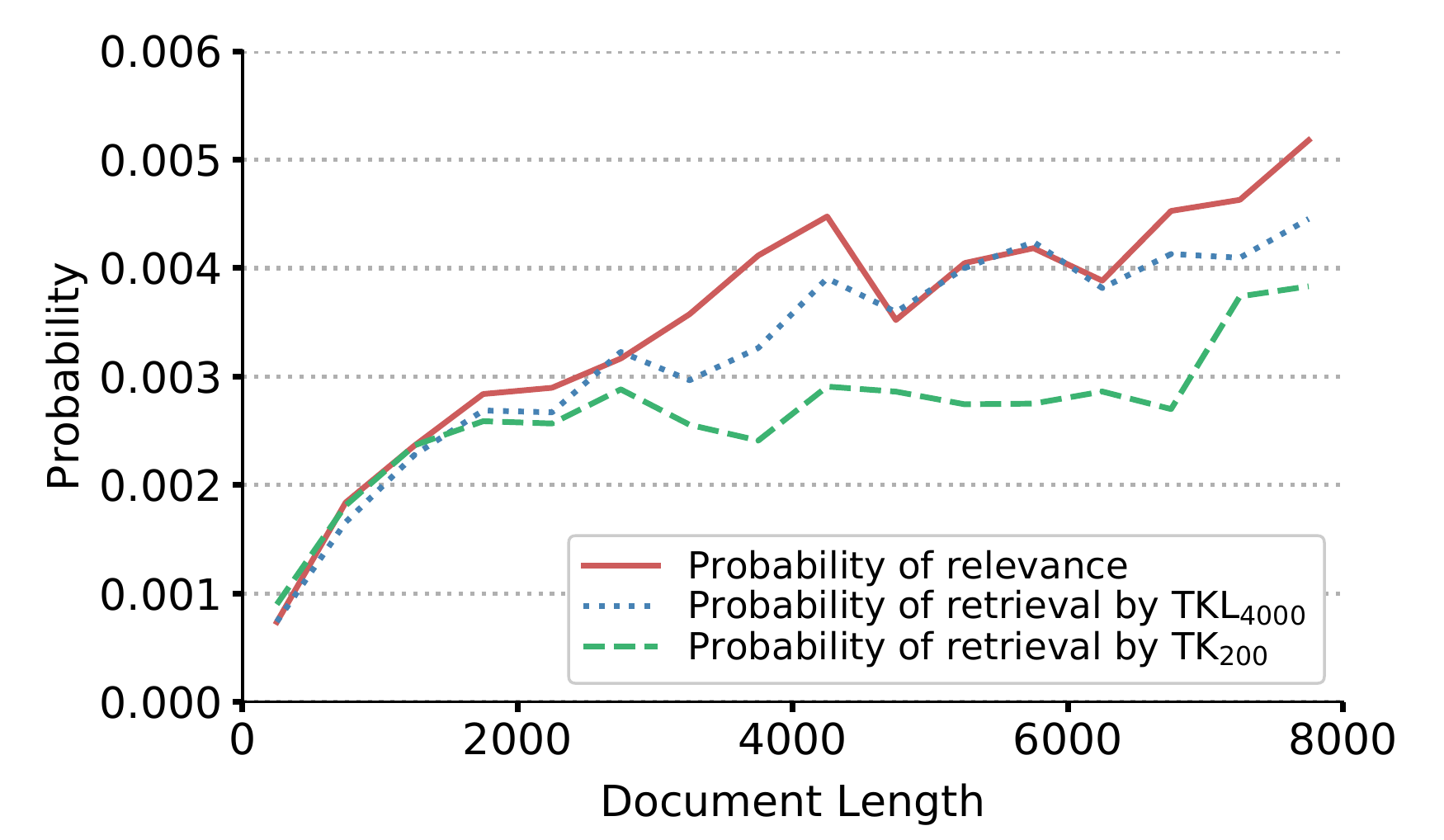}
    \vspace{-0.5cm}
    \caption{The probability of relevance and the probability of retrieval within top 30 ranks (using the baseline TK and our proposed TKL model) on the document retrieval task from TREC 2019 Deep Learning track.
    Longer documents have a disproportionately smaller probability of being retrieved by TK compared to their probability of relevance.
    TKL is less biased against retrieving longer documents.}
    \label{fig:doc_len_prob}
    \vspace{-0.5cm}
\label{fig:prob-rel-ret-len}
\end{figure}

A popular strategy involves considering only the first $n$ terms of a document.
However, this ignores any matches between the query and the remainder of the document which can result in poor retrieval quality, in particular for longer documents.
Fig.~\ref{fig:prob-rel-ret-len} illustrates the increasing gap between the \emph{probability of relevance} $P({rel} | l_i = {len}_{doc})$ and the \emph{probability of retrieval} $P({ret} | l_i = {len}_{doc})$ by the TK model as document length increases. 
A reasonable explanation may be that TK under retrieves longer documents because it inspects only the first $200$ terms of any document.

In this work, we propose a local self-attention which considers fixed-size moving windows over the document terms.
For each term, we only attend to the terms in the same window.
In case of non-overlapping windows, this reduces the time and memory complexity of self-attention over a sequence of length $L$ to $O(L \times l)$, where $l$ is the window size.
For $l \ll L$, this is a significant reduction in compute and memory requirements. In this work, we consider partially overlapping windows, with a slightly higher computation cost.

Another important challenge is to effectively aggregate the evidence from different parts of the document.
Towards that goal we propose a novel two-staged aggregation strategy:
\begin{enumerate*}[label=(\roman*)]
    \item A local aggregation, with a learned saturation function, within fixed-size windows, followed by
    \item a global selection of top-$t$ distinctly important regions of the document and corresponding signal aggregation.
\end{enumerate*}

We incorporate our proposed changes into a TK pooling model for long text (TKL), and answer the following research questions:
\begin{description}
    \item [R1] How does TKL compare to TK and other state-of-the-art retrieval methods?
    \item [R2] Does retrieval quality improve when TKL considers longer portions of documents?
    \item [R3] Is TKL more likely to retrieve longer documents than TK?
    \item [R4] What is the effect of learned saturation on retrieval quality?
    \item [R5] How often does TKL use different parts of the document?
\end{description}

All code used in this study is available at: \url{https://github.com/sebastian-hofstaetter/transformer-kernel-ranking}.

\section{Related Work}

\paragraph{Neural models for document retrieval}
Neural models have shown successful results in a number of IR tasks~\cite{guo2019survey,mitra2018introduction,Hofstaetter2019_sigir}.
\citet{Xiong2017} proposed a kernel pooling approach (KNRM) based on a bag-of-words representation of words.
This was further extended by \citet{Dai2018} to incorporate n-gram representations using convolutional architecture.
Several others \citep{Fan2018, jiang2019semantic} have highlighted important considerations for designing neural ranking models for documents that are distinct from dealing with passages and other short text. \citet{zamani2018snrm} have emphasized on efficiency in neural ranking models and introduced neural models for retrieving documents from a large corpus.
More recently, Transformer~\citep{devlin2018bert} based architectures have been employed to learn contextual representations which have led to bigger improvements~ \cite{nogueira2019passage, yan2020idst, Hofstaetter2020_ecai, padigela2019bertranking}.
\citet{yan2020idst} apply passage-level BERT-based relevance estimators to rank documents.
\citet{macavaney2019} use pretrained contextual embeddings, without fine-tuning, in downstream ranking models.

Classically, assessing relevance of documents based on relevant parts has been studied in many forms \cite{salton1993approaches,bendersky2008utilizing} and this study continues that exploration in the context of neural models.
Unlike \citep{macavaney2019, yan2020idst}, our proposed model is trained in a fully-supervised setting and only requires query-document relevance labels for training.

\paragraph{Efficient Transformers} 
\citet{al2019character} train a Transformer on the language modeling task by splitting long text into multiple segments.
\citet{dai2019transformer} extend that idea by incorporating a recurrence mechanism over the segments.
More recently, \citet{kitaev2020reformer} have proposed several techniques, including locality-sensitive hashing for self--attention, to scale the Transformer to longer text.
These techniques are orthogonal to the ideas presented in this paper and can be combined for additional efficiency gains.
\section{TKL Model}
\label{sec:model}

The TKL model adapts TK \cite{Hofstaetter2020_ecai} in two main aspects to enable document ranking: efficient attention \& scoring relevant regions.

We start by contextualizing query embeddings ($q_{1:n}$) in one window and document embeddings ($d_{1:m}$) in multiple windows of size $w$ and overlap them by $o$. Each window is contextualized by a multi-layered Transformer (TF) with highway connection and concatenated again, by removing overlapping vectors:
\begin{equation}
\begin{aligned}
\hat{q}_{1:n} &= \operatorname{TF}(q_{1:n}) \\
\hat{d}_{1:m} &= [\operatorname{TF}(d_{1:w+o})_{1:o};\operatorname{TF}(d_{w-o:2w+o})_{o:-o};...]
\end{aligned} 
\end{equation}

During batched processing of variable length padded documents, the independence of each window allows us to easily and efficiently skip segments that contain only padding, and pack the remaining windows to avoid unnecessary computations.

TKL transforms every individual term interaction with kernel-activations \cite{Xiong2017}, which splits similarities into activations based on the closeness to a certain range. Each kernel focuses on a fixed similarity range with center $\mu_{k}$ and width of $\sigma$. Each kernel results in a matrix $K \in \mathbb{R}^{|q| \times  |d|}$: 
\begin{equation}
K^{k}_{i,j} = \exp \left(- \frac{\left(\cos(\hat{q}_i,\hat{d}_j)-\mu_{k}\right)^{2}}{2 \sigma^{2}}\right)
\end{equation}

Now, TKL creates a relevance topography of a document, by sliding a saturation window across the interactions. This requires multiple steps. We start the process by computing the sum of document term interactions along dimension $j$ inside the sliding window region $r$ for each query term and kernel:
\begin{equation}
K^{r,k}_{i} = \sum_{j=1}^{j+r_{size}} K^{k}_{i,j}
\label{eq:kernel_doc_sum}
\end{equation}
\vspace{-0.2cm}

Instead of a fixed log saturation (as used by previous kernel-pooling models \cite{Xiong2017,Dai2018,Hofstaetter2020_ecai}) we learn the shape of our non-linear saturation function:
\begin{equation}
\widehat{K^{r,k}_{i}} = a_i * \left( K^{k}_{i} \right)^{1/b_i} - c_i
\label{eq:kernel_doc_activation}
\end{equation}
where $a_i$,$b_i$,$c_i$ are conditioned on a query term salience embedding $e_i$ and the region token count $c_{len}$ (to not disadvantage regions containing padding):
\begin{equation}
\begin{aligned} 
a_i = [ReLU(e_i); c_{len}] * W_a + b_a \\
b_i = [ReLU(e_i); c_{len}] * W_b + b_b \\
c_i = [ReLU(e_i); c_{len}] * W_c + b_c \\
\end{aligned} 
\label{eq:kernel_doc_activation_details}
\end{equation}

We initialize $b_*$ with 100, so that the training starts with a log-approximation and $e_i$ with the Inverse Document Frequency of the collection per term.
After the saturation, we sum query dimensions, and weight kernel bins, to receive a relevance topography over the document:
\begin{equation}\label{eq:rsv}
s^r = \bigg(\sum_{i=1}^{|q|}\widehat{K^{r, k}_{i}} \bigg) W_k
\end{equation}

Finally, \textit{top-local-max} takes the top-t local maxima and their $f$ immediate neighbors, by selecting the 1 to $f$ left and right values of the maxima. By that $W_s$ may learn a combination between the peak and the slope of the topography of the most relevant regions:
\begin{equation}\label{eq:rsv}
s = \text{top-local-max}_{t,f}\bigg(s^r\bigg) W_s
\end{equation}

We define local as the saturation region size $r$, so that we do not count term matches twice. The position of the regions can easily be extracted with the final output score, enabling the user interface to highlight these regions. Furthermore, it allows us to analyze the TKL model as we do in Section \ref{sec:results}.

\begin{table*}[t!]
    \centering
    \caption{Effectiveness and efficiency results for both query sets. For the stat. significance ${a-f}$ includes ${abcdef}$.}
    \label{tab:all_results}
    \vspace{-0.3cm}
    \setlength\tabcolsep{4pt}
    \begin{tabular}{cll!{\color{lightgray}\vrule}rrr!{\color{lightgray}\vrule}rrr!{\color{lightgray}\vrule}r}
       \toprule
       \multirow{2}{*}{Sig.} & \multirow{2}{*}{\textbf{Model}}&
       \textbf{Max Doc.}   &
       \multicolumn{3}{c!{\color{lightgray}\vrule}}{\textbf{TREC DL Track 2019}}&
       \multicolumn{3}{c!{\color{lightgray}\vrule}}{\textbf{TREC 2019 Dev - Sparse Labels}}&
        \textbf{Average}\\
       && \textbf{Length}   & nDCG@10 & MRR@10 & MAP@100 & nDCG@10 & MRR@10 & MAP@100  & \textbf{Docs./ms}\\
        \midrule
        \multicolumn{6}{l}{\textbf{Baselines}} \\
        $a$ & BM25     & - &   0.488 & 0.815 & 0.234 & 0.311 & 0.252 & $^{e}$0.265 &   - \\
        $b$ & MatchPyramid \citep{Pang2016}     & 200 & $^{e}$0.567 & $^{e}$0.903 & $^{e}$0.232 & $^{ae}$0.344 & $^{ae}$0.286 & $^{e}$0.288  & 27 \\
        $c$ & PACRR \citep{hui2017pacrr}           & 200 & $^{ae}$0.606 & 0.860 & $^{e}$0.228 & $^{ae}$0.344 & $^{ae}$0.283 & $^{e}$0.286  & 22 \\
        $d$ & CO-PACRR \citep{hui2018co}         & 200 & $^{e}$0.550 & $^{e}$0.895 & $^{e}$0.231 & $^{ae}$0.345 & $^{ae}$0.284 & $^{e}$0.288  & 14 \\
        $e$ & KNRM \citep{Xiong2017}            & 200 & 0.496 & 0.771 & 0.214 & $^{a}$0.323 & $^{a}$0.261 & 0.264  & {49} \\
        $f$ & CONV-KNRM \citep{Dai2018}       & 200 & $^{e}$0.565 & $^{e}$0.903 & $^{e}$0.241 & $^{ae}$0.345 & $^{ae}$0.283 & $^{e}$0.287   & 10 \\

        $g$ & BERT$_\textbf{[CLS]}$ \citep{nogueira2019passage} & 200 &  $^{a-f}$0.642 & $^{ace}$0.944 & $^{e}$0.257 & $^{a-fhij}$\textbf{0.417} & $^{a-fhij}$\textbf{0.352} & $^{a-fhij}$\textbf{0.358}   & 0.1 \\
        $h$ & TK \citep{Hofstaetter2020_ecai}                   & 200 &  $^{e}$0.594 & $^{e}$0.903 & $^{cde}$0.252 & $^{a-f}$0.375 & $^{a-f}$0.312 & $^{a-f}$0.318 & 4 \\
        \midrule
        \multicolumn{6}{l}{\textbf{Best single BERT-based official TREC 2019 runs}} \\
        - & ucas\_runid1           & n/a  &  0.644 & 0.911 & 0.264 & - & - & - & <0.1 \\
        - & bm25\_marcomb \cite{yilmaz2019} & n/a &  0.640 & 0.913 & \textbf{0.323}  & - & - & - & <0.1 \\ 
        \midrule

        \multicolumn{6}{l}{\textbf{Our proposed models}} \\
        $i$ & TKL & 2,000 & $^{a-fh}$0.634 & $^{e}$0.915 & $^{cdef}$0.264 & $^{a-fhj}$0.403 & $^{a-fhj}$0.338 & $^{a-fh}$0.345 & 1.1 \\
        $j$ & TKL & 4,000  & $^{abdef}$\textbf{0.644} & $^{ace}$\textbf{0.957} & $^{cdei}$0.277 & $^{a-fh}$0.396 & $^{a-fh}$0.329 & $^{a-fh}$0.336  & 0.9 \\

        \bottomrule
    \end{tabular}
    \vspace{-0.3cm}
\end{table*}

\section{Experiment Design}
\label{sec:exp}

We utilize the recent TREC Deep Learning track dataset \cite{trec2019overview} for document retrieval, derived from MS MARCO \cite{msmarco16}. It contains 3.21 million documents. The median word count is 804, with the 90\textsuperscript{th} percentile including 3267 words. 
For evaluation we use two differently created test sets: Dev sparse judgements on 5193 queries (only 1 relevant-judged document per query) and high-qualitative dense judgements (deep pooling) from the 2019 Deep Learning track on 43 queries (on average 378 relevant documents per query).

To have a fair comparison, we utilize the provided initial ranking of top 100 documents returned by BM25 for all models, except for the bm24marcob baseline runs that uses a stronger first stage retrieval. We conduct statistical significance tests with a Wilcoxon signed-rank test with $p<0.05$. 

For the parameter settings of the baselines, we followed the settings from Hofstätter et al. \cite{Hofstaetter2020_ecai}. All models except BERT use 300 dimensional GloVe embeddings.
TK and TKL use 2 Transformer layers with 10 attention heads. TKL uses a chunk width $w$ of 40, overlapping $o$ of 10, region size $r$ of 30, and weights the top $3$ local-maxima with $2$ neighboring near-maxima to form the final score. For kernel-activation we use the default of $11$ kernels from $-1$ to $+1$ and standard deviation of $0.1$. 
We train with a batch size of 32 and the Adam optimizer with a learning rate of $10^{-4}$ for representation learning, $10^{-3}$ for other network components. Our early stopping is based on the best nDCG@10 validation value. Our efficiency measurements are based on NVIDIA GTX 1080 GPUs.
\vspace{-0.2cm}
\section{Results}
\label{sec:results}

\noindent
\textbf{R1.} \emph{How does TKL compare to TK and other SOTA retrieval methods?}

Table~\ref{tab:all_results} compares TKL to several baselines -- \incl TK and other SOTA neural models on the TREC Deep Learning document ranking task.
Our main result in this paper is that the proposed TKL model achieves significant improvements over the TK baseline.
TKL achieves comparable performance to BERT models in case of complete judgements, which are more reliable than the sparse labels of the Dev set. 

The TKL model operating on 10 to 20 times more tokens, is more effective than the BERT model only considering 200 tokens. Note that the high GPU memory requirements of BERT does allow us to extend to many more terms.

Table~\ref{tab:all_results} also contains the results of two successful runs from the TREC DL Track 2019, i.e., ucas\_runid1 and bm25\_marcob.\footnote{Note that we do not consider ensemble models for fair comparison.} Both of these models used BERT for document retrieval. In other words, they chunk the documents and produce scores per passages and combine the scores for document retrieval. Their results are also comparable to those obtained by the TKL model, in terms of nDCG@10. The MRR obtained by TKL is higher than those obtained by these two models. The bm25\_marcob model achieves higher MAP, however this is due to the stronger first stage retrieval used by this model. In other words, all the models except for bm25\_marcob re-rank the top 100 documents provided by the track organizers, while bm25\_marcob uses a full ranking strategy by using a stronger first stage retrieval model.

\smallskip
\noindent
\textbf{R2.} \emph{Does retrieval quality improve when TKL considers longer portions of documents?}

\begin{figure}
    \centering
    \includegraphics[width=0.4\textwidth,clip, trim=0.5cm 0.1cm 0cm 0cm ]{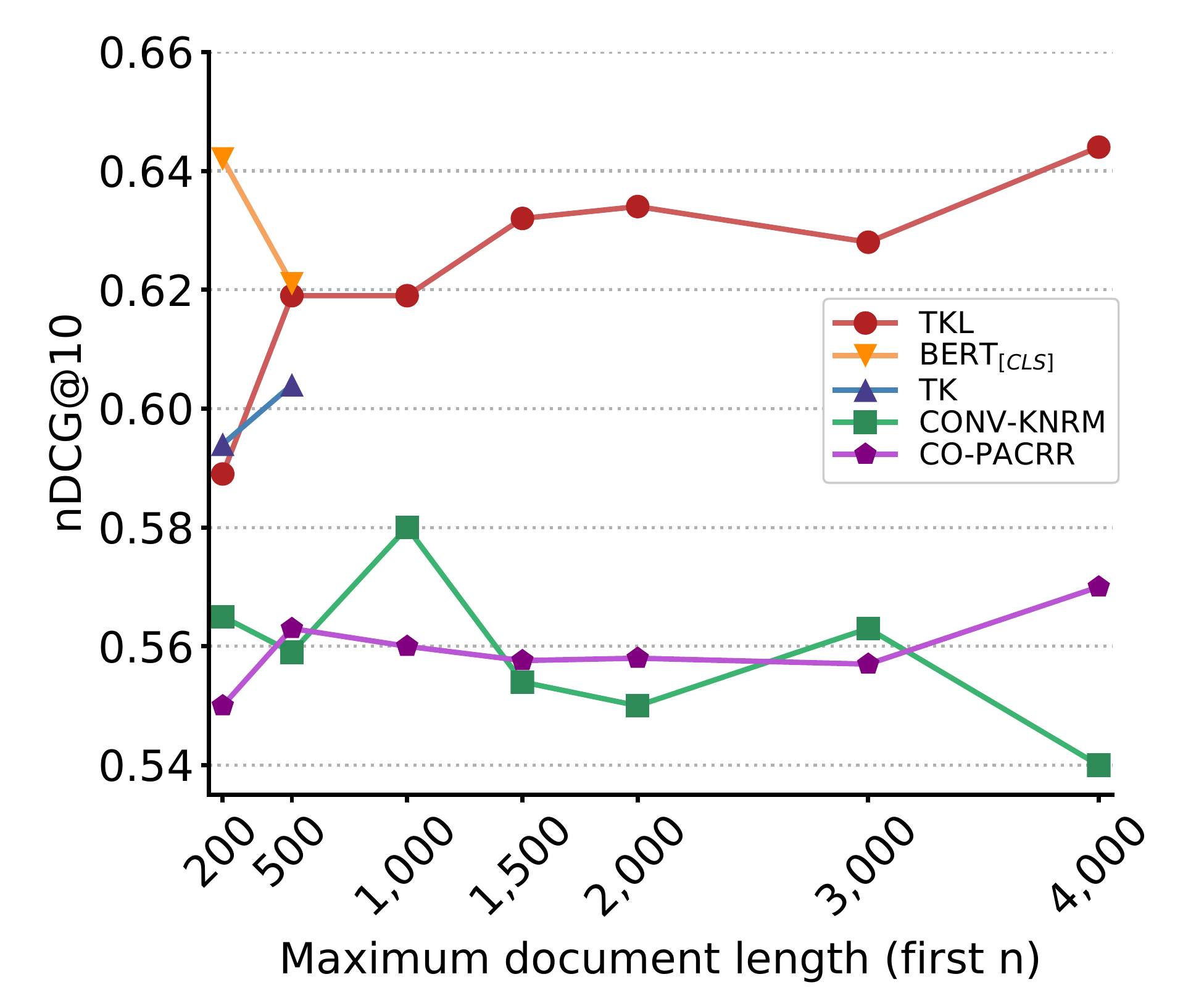}
    \vspace{-0.2cm}
    \caption{TREC-2019 results based on the document length.}
    \label{fig:per_doc_len}
    \vspace{-0.3cm}
\end{figure}

Fig.~\ref{fig:per_doc_len} shows the nDCG@10 results of our models trained and evaluated on different maximum document lengths.
We observe the strength of the TKL model: The longer the input document, the better the results. The document-wide kernel-pooling saturation used by CONV-KNRM does not clearly benefit from longer text.

\smallskip
\noindent
\textbf{R3.} \emph{Is TKL more likely to retrieve longer documents than TK?}

Figure \ref{fig:doc_len_prob} shows that TKL is more likely than TK to retrieve long relevant documents. While for short documents, there is no difference between TK and TKL, the improvement of TKL is distinct for longer documents, as its probability to retrieve longer documents is closer to the probability of retrieving relevant documents.

\smallskip
\noindent
\textbf{R4.} \emph{What is the effect of learned saturation on retrieval quality?}

\begin{table}[t!]
    \centering
    \caption{Ablation study of different saturation functions using 2,000 words per document.}
    \label{tab:sat_ablation_results}
    \vspace{-0.3cm}
    \setlength\tabcolsep{3pt}
    \begin{tabular}{cl!{\color{lightgray}\vrule}rr!{\color{lightgray}\vrule}rr}
       \toprule
       \multirow{2}{*}{Sig.} & \textbf{TKL}&
       \multicolumn{2}{c!{\color{lightgray}\vrule}}{\textbf{TREC DL 2019}}&
        \multicolumn{2}{c}{\textbf{TREC DL Dev-Sparse}} \\
       &\textbf{Saturation} & nDCG@10 & MAP & nDCG@10 &  MAP  \\
        \midrule

        $a$ & \textit{\textbf{Linear}}    & 0.570 & 0.237 & 0.366 & 0.308  \\
        $b$ & \textit{\textbf{Log}}       & $^{a}$0.618 & $^{a}$\textbf{0.266} & $^{a}$0.400 & $^{a}$0.341  \\
        $c$ & \textit{\textbf{Embedding}} & \textbf{0.634} & $^{a}$0.264 & $^{ab}$\textbf{0.403} &  $^{ab}$\textbf{0.345}  \\

        \bottomrule
    \end{tabular}
    \vspace{-0.4cm}
\end{table}

Table \ref{tab:sat_ablation_results} shows an ablation study of different saturation functions (as defined in Eq. \eqref{eq:kernel_doc_activation}). It is clear, that the saturation requires a non-linear shape, as the linear version (with $b=1$) suffers strongly in comparison to the others. Furthermore, our novel query salience conditioned function outperforms the fixed log function used in previous kernel-pooling approaches, except for MAP on the TREC DL 2019 dataset.

\smallskip
\noindent
\textbf{R5.} \emph{How often does TKL attend to different parts of the document?}

\begin{figure}
    \centering
    \includegraphics[width=0.478\textwidth,clip, trim=0.5cm 0.2cm 0cm 0cm ]{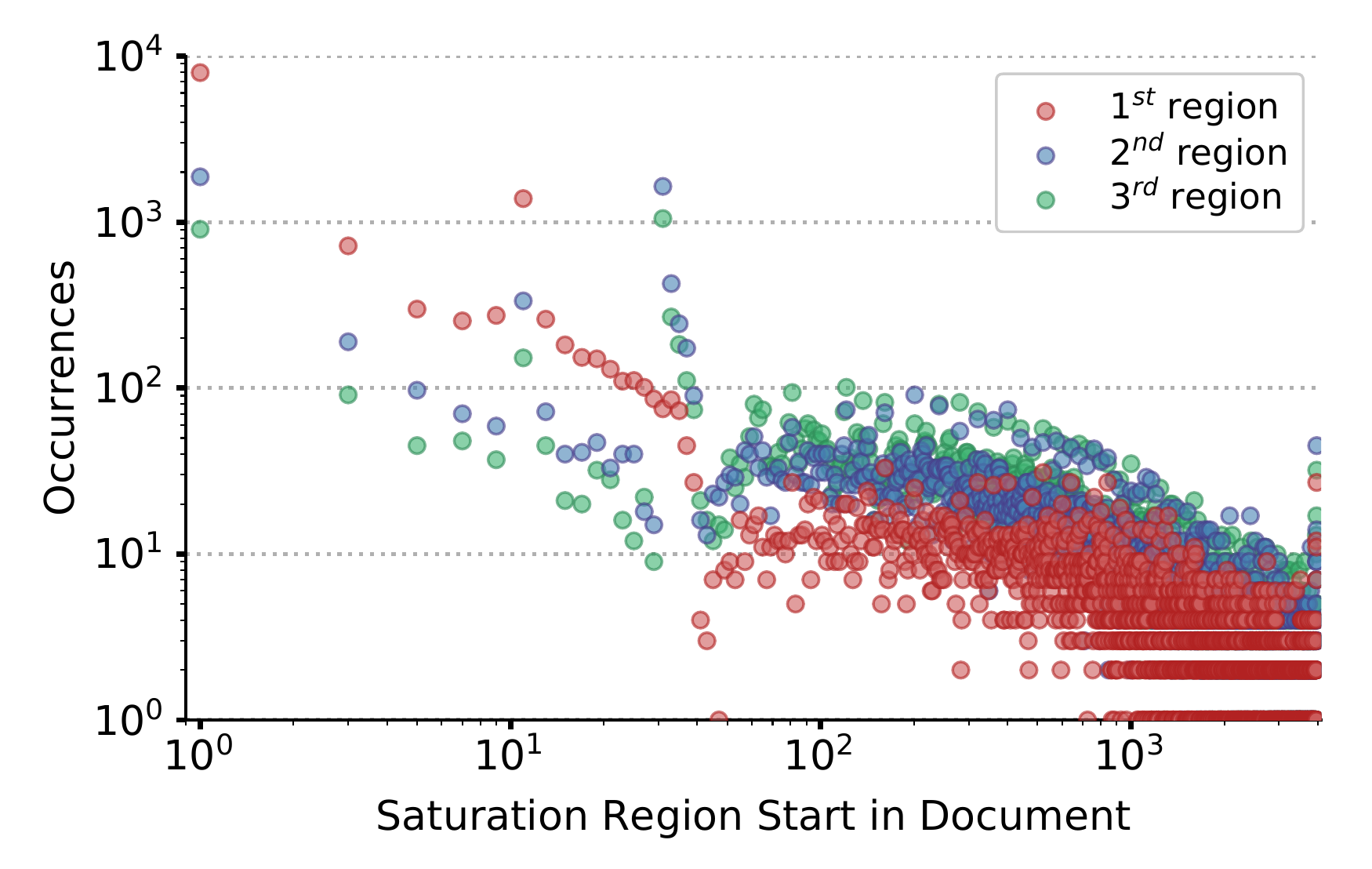}
    \vspace{-0.8cm}
    \caption{Positions of the top-3 regions in TREC-2019 of TKL using a log-log scale.}
    \label{fig:topk_regions}
    \vspace{-0.5cm}
\end{figure}

We show the distribution of the top-3 relevant regions in Figure \ref{fig:topk_regions}. While we can clearly see a focus on the beginning of the document, we observe a sizeable amount of relevant regions after 500 words. The most relevant region occur 24.5 \%, the second 41.5 \%, and the third 46.4 \% of the time after the 500$^{th}$ word. Even though our baselines achieve acceptable results by only looking at the start of a document, this result in Figure \ref{fig:topk_regions} shows that TKL learns to detect relevant regions in every part of a document. 

\smallskip

\section{Conclusion}
\label{sec:conclusion}
In this work we proposed a solution to apply Transformers to full document re-ranking. Our TKL model efficiently contextualizes overlapping windows, which allows us to pack padded documents easily. Furthermore, we proposed a novel saturation function, conditioned on query term salience, to slide over a document and detect the top distinct relevant regions in the document. Our experiments on the TREC Deep Learning datasets showed that TKL takes advantage of the increased input. We observed improving performance as more input tokens are fed to the model. Therefore, TKL provides effective performance with high efficiency while using thousands of terms from the documents.

\bibliographystyle{ACM-Reference-Format}
\bibliography{sigproc} 
\end{document}